\documentclass[superscriptaddress,amsmath,amssymb]{revtex4}

\usepackage{amsmath}
\usepackage{amssymb}
\usepackage{graphicx}
\usepackage{citesort}

\newcommand{\vect}[1]{{\bf #1}}
\newcommand{\uvect}[1]{{\bf \hat{#1}}}
\newcommand{\order}[1]{\mathcal{O}\left( #1 \right)}

\newlength{\gxlen} 
\newlength{\hgxlen} 
\newlength{\boxlen} 
\newlength{\pboxlen} 

\newcommand{\fref}[1]{Fig.~\ref{fig:#1}}
\newcommand{\Fref}[1]{Fig.~\ref{fig:#1}}
\newcommand{\sref}[1]{section~\ref{sec:#1}}
\newcommand{\Sref}[1]{Section~\ref{sec:#1}}
\newcommand{\lqq}{``}
\newcommand{\rqq}{''}

\begin{document}

\title{Event-Driven Molecular Dynamics in Parallel}          



\author{S. Miller}
\affiliation{
  Institut f\"ur Computeranwendnungen~1, 
  Universit\"at~Stuttgart, Pfaffenwaldring~27, D-70569~Stuttgart, 
  Germany
  } 
\author{S. Luding}
\affiliation{
  Institut f\"ur Computeranwendnungen~1, 
  Universit\"at~Stuttgart, Pfaffenwaldring~27, D-70569~Stuttgart, 
  Germany
  } 
\affiliation{
  Particle Technology, DelftChemTech, TU~Delft, 
  Julianalaan~136, 2628~BL~Delft, The~Netherlands
  } 

\date{\today}

\begin{abstract}
Although event-driven algorithms have been shown to be far more efficient 
than time-driven methods such as conventional molecular dynamics, 
they have not become as popular.
The main obstacle seems to be the difficulty of parallelizing 
event-driven molecular dynamics.
Several basic ideas have been discussed in recent years, 
but to our knowledge no complete implementation has been published yet.
In this paper we present a parallel event-driven algorithm 
including dynamic load-balancing, which can be easily implemented 
on any computer architecture. 
To simplify matters our explanations refer to a basic multi-particle system
of hard spheres, 
but can be extended easily to a wide variety of possible models.
\end{abstract}


\maketitle


\section{Introduction}

Event-driven molecular dynamics is an effective algorithm for the simulation 
of many-component systems, which evolve independently, 
except for discrete asynchronous instantaneous interactions.
As an example we will discuss a system consisting of $N$ hard spheres 
in a box with periodic boundary conditions,
but the algorithm can be extended to particles with different shapes and 
interacting via any piecewise constant potential 
or to completely different problems as well
\cite{lubachevsky00}.

Event-driven molecular dynamics processes a series of 
events asynchronously one after another.
A straight-forward but simplistic approach \cite{alder59} 
updates all particles at each event; many advanced versions 
are available in text-books \cite{allen87,rapaport95}.
We base our algorithm on the sophisticated algorithm as 
presented by \cite{lubachevsky91},
which updates only those particles involved in an event.
It has been succesfully applied 
to many different problems, among them 
granular gases \cite{luding99,luding02c}, polymer chains \cite{smith97}, 
tethered membranes 
\cite{luding02c,sm:memb}, protein folding \cite{dokholyan98},
and battlefield models \cite{nicol88}.

In many physical systems the duration of the interaction of the components,
e.\,g.\ the collision of two particles, is negligible compared to the time 
between these interactions.
The simulation of such systems with traditional time-driven molecular dynamics 
is highly inefficient. 
Instead, it is straight-forward to consider the interactions as 
instanteneous events and solve the problem with an event-driven algorithm.

One reason why event-driven molecular dynamics has not become 
as popular as conventional time-driven molecular dynamics is the fact that
parallelization is extremely complicated.
The paradoxical task is to algorithmically parallelize physically 
non-parallel dynamics.
Nevertheless some ideas and basic considerations about the parallelization 
have been proposed in \cite{lubachevsky92}.
They are especially suited for shared memory computers,
but can be transferred to distributed memory architectures as well 
\cite{marin97}.
Apart from those ideas no full and general implementation 
of a parallelized algorithm including load-balancing has been published yet,
to our knowledge.
In this paper we present an algorithm, which is based on those ideas,
but is enhanced and completed at several points. 
It can be implemented with generic tools such as MPI, 
and therefore it is suited for shared and distributed memory 
architectures alike.

In \sref{ed} we explain the details of the implementation of 
event-driven molecular dynamics.
The parallelization of the algorithm is presented in \sref{parallel} and
a summary is given in \sref{summary}. 

\section{Event-Driven Molecular Dynamics}
\label{sec:ed}

\Sref{outline} gives an outline of the main routine of the algorithm, which 
is composed of 4 steps (see \fref{serial}).
Then the most important data structures are introduced in \sref{data} 
and hereby step 1 and 4 are treated.
Step 2 and 3 are discussed in sections \ref{sec:update} 
and \ref{sec:eventcalculation}, respectively. 
Finally, \sref{optimum} deals with performance issues.

\subsection{Outline}
\label{sec:outline}

Event-driven molecular dynamics processes a series of discrete events.
In a system of hard spheres events typically refer 
to instantaneous collisions, involving exactly two particles.
Only these two particles are processed; 
the state of the other particles remains unchanged.
So the state information of most particles refers not to the current time, 
but to a different point of time in the past.

Event processing includes a state update for the concerned particles 
(see \sref{update}) 
and the calculation of the next future events for those particles
(see \sref{eventcalculation}).
An outline of the algorithm can be seen in \fref{serial}.

\begin{figure}
  \framebox[\boxlen]{
    \parbox{\pboxlen}{
      \begin{flushleft}
      {\bf event processing loop:}
      \begin{enumerate}
      \item get next event from priority queue $\to$ current time
      \item update states of both particles to current time
      \item calculate new future events for both particles 
      \item schedule next event for both particles in priority queue
      \item goto 1      
      \end{enumerate}
      \end{flushleft}
      }
    }
  \caption{Outline of the main routine of the algorithm}
  \label{fig:serial}
\end{figure}

The {\em serial} algorithm only adds to this event processing loop 
an initialization step at the beginning and periodic data output.
The {\em parallel} algorithm needs several additional routines, which are
described in \sref{parallel}.

\subsection{Data Organization}
\label{sec:data}
\label{sec:eventlist}
\label{sec:pqueue}

The algorithm maintains two data fields, {\em particle states} 
and {\em event list}, which contain exactly one entry per particle 
(see \fref{data}).
The former refers to the past of a particle, the latter to its future
($\leadsto \forall i : \mbox{state time }t_0(i) \le 
\mbox{current time} \le \mbox{event time }t_{\rm ev}(i)$).

\begin{figure}
  \framebox[\boxlen]{
    \parbox{\pboxlen}{
        \begin{align*}
        &\mbox{particle states} &: N \times &\left\{ 
          t_0, \vect{r}(t_0), \vect{v}(t_0), \dots,   
          \mbox{counter}, \mbox{cell}, \mbox{id} 
        \right\}\\
        &\mbox{event list} &: N \times &\left\{ 
          t_{\rm ev}, \mbox{type}, \mbox{partner}, 
          \mbox{counter(partner)} 
        \right\}
      \end{align*}
      }
    }
  \caption{Data structures for $N$ particles}
  \label{fig:data}
\end{figure}

A {\em particle state} consists of the physical state of a particle, 
such as position, velocity, etc.\ immediately after the most recently 
processed event of that particle, the point of time of that event, 
an event counter (see \sref{update}), a cell number (see \sref{cells}), and
a global particle number, which is needed to identify particles 
on different processes (see \sref{parallel}).

The {\em event list} associates with every particle an event in the future.
The data units of the {\em event list} consist of event time, event type, 
event partner, and a copy of the event counter of the partner.

All the events are scheduled in a priority queue
which is usually implemented as an implicit heap tree
\cite{lubachevsky91,lubachevsky92,knuth68}, 
but other data structures with similar characteristics are possible, 
too 
\cite{marin95,knuth68}.
A heap tree is a binary tree with the following ordering rule:
Each parental node has higher priority than its children; 
the children themselves are not ordered.
So the root node always has highest priority, 
i.\,e.\ the earliest event time.
To get the next event from the priority queue takes computational costs of
$\order{1}$.
Insertion of a new event in the priority queue is done with costs of 
$\order{\log N}$. 

This data organization is more efficient compared 
to the algorithm in \cite{lubachevsky91,lubachevsky92}, 
since no double buffering is needed here, however,
the basic ideas of \cite{lubachevsky91,lubachevsky92} are still valid.

\subsection{State Update}
\label{sec:update}

When a collision between two particles is processed, the states of 
these particles are updated from a point of time 
in the past to the current simulation time.
First, the positions and velocities of the particles 
immediately before the collision
can be derived by inserting the old state in the equation of motion.
As a result of this first step the particles are touching each other.
Then the interaction between the particles takes place and yields 
new particle velocities.
Now, the event counter is increased, the state time is updated 
to the current simulation time, and the values of 
the positions and velocities immediately after the collision are stored 
in the {\em state} array.

A typical collision rule for hard spheres \cite{luding01d,luding02b} 
looks like 
\begin{align*}
\vect{v'}_{1/2} &= \vect{v}_{1/2} \mp \frac{1+r}{2} 
\left( \uvect{k} \cdot ( \vect{v}_1- \vect{v}_2 ) \right) \uvect{k} \; ,
\end{align*}
where primes indicate the velocities $\vect{v}$ after the collision, 
$\uvect{k}$ is a unit vector pointing along the line of centers 
from particle 1 to particle 2, and $r$ denotes the restitution coefficient.
For the performance tests in the subsequent chapters 
we have used the simplest case without dissipation ($r=1$).
\footnote{Soft spheres can be approximated via particles consisting of 
concentric shells with piecewise constant interaction potentials.}

If the event partner has undergone collisions with other particles 
between the scheduling and the processing of the event,
it becomes invalid.
Validity of event partners can be checked by comparing 
the event counter of the partner to the copy in the event list.
If they do not match, the partner has collided with other particles 
in the meantime, and the scheduled collision is no longer valid.
Then event processing only refers to one particle and
the state update described above only consists of the particle motion; 
no collision is performed.
After that the algorithm continues in the normal way, i.\,e.\ the next event 
for the particle will be calculated and scheduled in the priority queue.

Note that the algorithm in \cite{lubachevsky91,lubachevsky92} uses 
a different strategy to detect invalid event partners:
The algorithm checks at each collision if an event partner becomes 
invalid and if so marks this partner.
This strategy is less efficient, because sometimes the same partner 
is invalidated several times.
Besides, in the parallel algorithm additional communication 
between different processes might be necessary.

\subsection{Event Calculation and Linked Cell Structure}
\label{sec:eventcalculation}
\label{sec:cells}

When an event has been processed, new events for the particles involved in that
event have to be calculated.
In simulations of hard spheres this means the calculation of 
possible future collisions.
If the particles move on ballistic trajectories 
\begin{align*}
\vect{r}_i(t) = \vect{r}_i(t_0)+\vect{v}_i(t_0)\,(t-t_0)
+\frac{1}{2}\vect{g}\,(t-t_0)^2 \;,
\end{align*}
two particles 1 and 2 will 
collide at time $t_{12}$:
\begin{align*}
  t_{12} &= t_0 + \left( -\vect{r}_{12} \cdot \vect{v}_{12}
    -\sqrt{(\vect{r}_{12} \cdot \vect{v}_{12})^2 
      - \left( r_{12}^2-(R_1+R_2)^2 \right) v_{12}^2} 
  \right) / v_{12}^2 \;,
\end{align*}
where $\vect{v}_{12}=\vect{v}_2(t_0)-\vect{v}_1(t_0)$ and 
$\vect{r}_{12}=\vect{r}_2(t_0)-\vect{r}_1(t_0)$ are the relative velocities and
positions of the particles at time $t_0$, 
and $R_i$ are the radii of the particles.
If $t_{12}$ is imaginary or smaller than $t_0$, the particles will not collide.

If the algorithm would have to check for collisions with all other particles, 
performance would be very poor for large numbers $N$ of particles.
So we divide simulation space in $C$ cells with equal sides.
Each particle belongs to the cell in which its center lies.
If the cell size is larger than the maximal interaction length 
of the particles, i.\,e.\ the particle diameter in the case of hard spheres, 
event calculation has to check for possible collisions of two particles only 
if they belong to the same cell or if their cells are neighbors.
(One square cell has 8 neighbors in 2D and a cube has 26 neighbors in 3D.)

However, the algorithm has to keep track of cell changes.
So additional events, namely cell changes, come into play.
They are treated just in the same way as the collision events; 
only the collision partner is the boundary of a cell.
The difference with a collision event is that only one particle 
is involved in a cell change, 
and the velocity of the particle does not change at event time.
Cell changes at the boundary of the simulation space require that
the position vector jumps to the opposite side of the simulation volume 
due to the periodic boundary conditions.

\subsection{Optimal Cell Numbers}
\label{sec:optimum}

On average there are $3^D N/C - 1$ particles in the neighborhood 
of each particle.
So in the limit $C \ll N$ this number becomes very large 
and many possible events have to be calculated after each collision.
On the other hand, if $C \gg N$, then each particle has to cross 
many cell boundaries between a collision of two particles, and thus 
more events have to be treated to complete the simulation.
These two contributions compete with each other, the first one is proportional 
to $( C/N )^{-1}$ and the second one is proportional to 
$(C/N)^{1/D}$.
\Fref{cellopt} (left) shows a broad minimum 
of the simulation time for low densities at the optimal cell number 
$C/N \approx 1.5$ in 2D and 
$C/N \approx 8$ in 3D.
So the program can choose an optimal cell number before the calculation 
starts.
Note that in the high density limit, especially in 3D, 
the optimal cell number cannot be reached, 
because the size of the cells has to be larger than the particle diameter
(see \fref{cellopt} (right)).
So, in this case $C$ should simply be chosen as big as possible.

\begin{figure}
  \begin{center}
    \includegraphics*[width=\hgxlen]{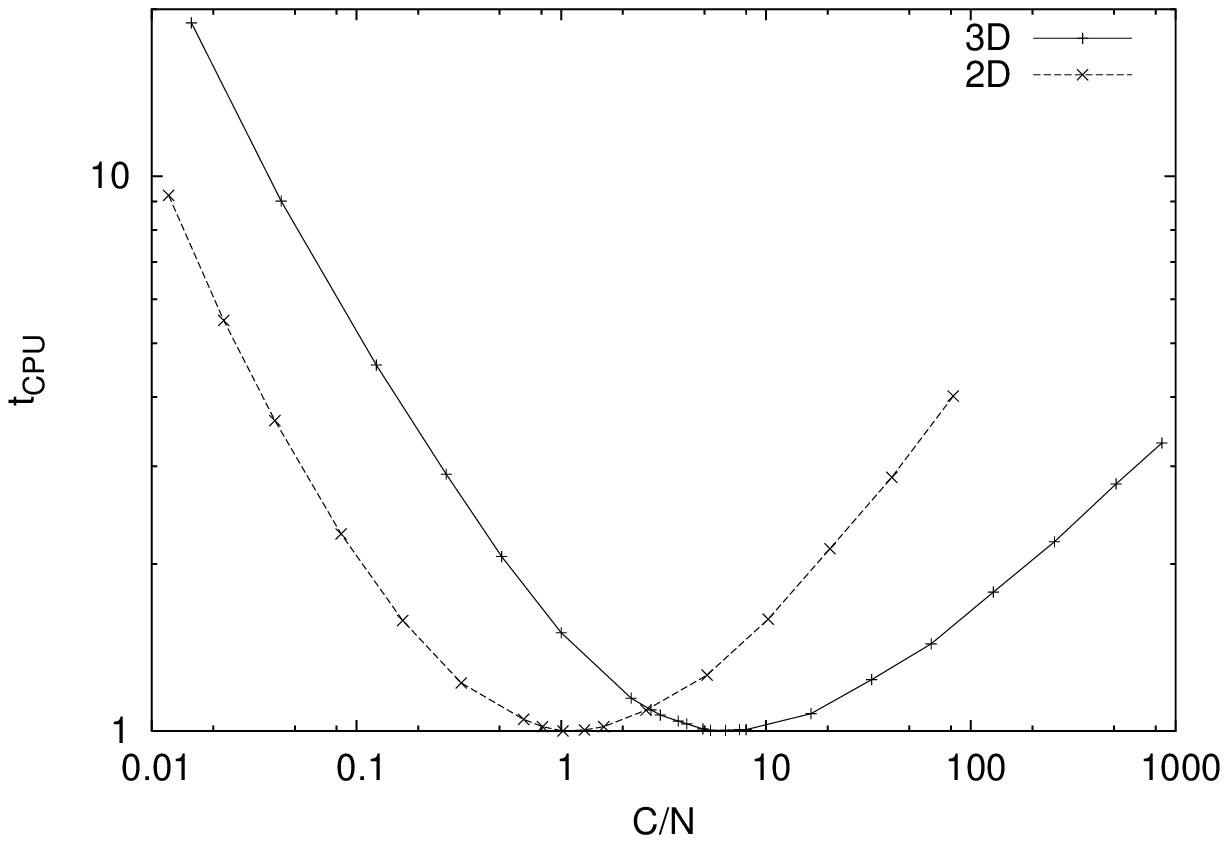}\hfill
    \includegraphics*[width=\hgxlen]{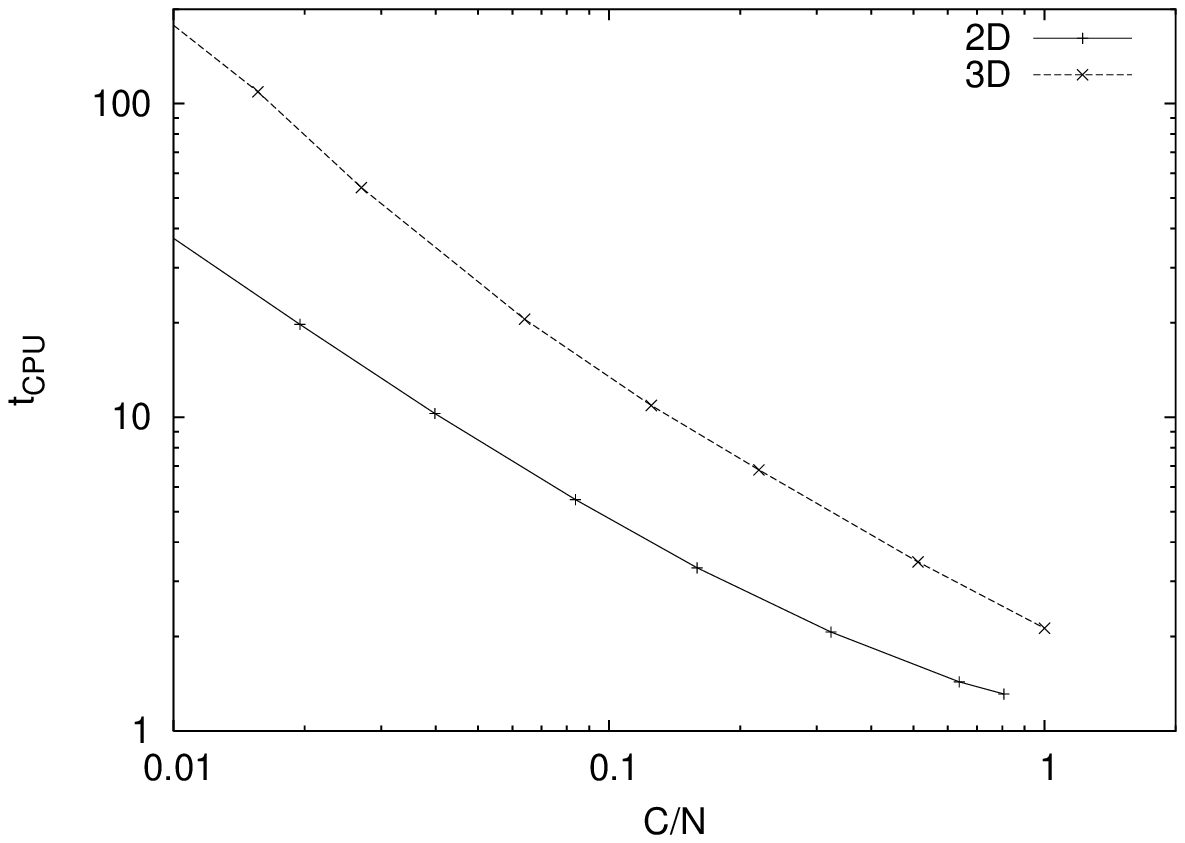}\\*
  \end{center}
    \caption{Simulation time $t_{\rm CPU}$ (arbitrary units) 
      plotted against the number $C$ of cells in 2D and 3D 
      with the serial algorithm.
      Used are $N=10000$ particles (2D) and $N=8000$ particles (3D).
      The left figure shows low densities (volume fraction $\nu=0.008$ (2D) 
      and $\nu=0.0005$ (3D), the right figure shows high densities 
      ($\nu=0.8$ (2D) and $\nu=0.5$ (3D)).}
    \label{fig:cellopt}
\end{figure}

\section{The Parallel Algorithm}
\label{sec:parallel}

In \sref{domain} we demonstrate how parallelization can be achieved 
via domain decomposition and dynamic load-balancing.
Then, in \sref{causalorder} we point out the difficulties that arise 
in parallelizing this algorithm and
the necessity of state saving and error recovery (see \sref{statesave}).
In \sref{paroutline} the parallel algorithm is explained in detail.
Finally, some performance issues are discussed in \sref{performance}. 

\subsection{Domain Decomposition and Dynamic Load-Balancing}
\label{sec:domain}
\label{sec:loadbalancing}

Parallelization is achieved via domain decomposition.
Each cell is affiliated with a process, but this affiliation can 
change during the simulation if the load of the processes has to be rebalanced.

In order to realize maximal performance, the computational load 
should be distributed equally among the processes.
If inhomogenities exist from the very beginning or emerge 
during the simulation, dynamic load-balancing becomes important.
A typical example is cluster formation in granular gases.
\cite{luding99,luding02}

Domain decomposition should thus take into account the following points:
Firstly, the load of the processes should be distributed homogeneously. 
Secondly, in order to minimize process communication, 
the border area of each process domain should be as small as possible, 
which implies that the ideal process domains are squares or cubes.
Thirdly, a simple and fast function that assigns a cell to a process 
and vice versa is required.

We meet these demands in the following way:
Cell numbers are assigned to cell coordinates 
in a tree-like structure, see a small 2D-example in \fref{celllayout}. 
(Note that a realistic example has thousands or millions of cells.)
These cell numbers are obtained by interleaving all the bits $1$ to $k$ of 
the cell coordinates: $z[k],y[k],x[k],\dots,z[1],y[1],x[1]$.
The result is a cell number in the range $0$ to $2^{kD}-1$. 
(For convenience these numbers are increased by one in \fref{celllayout}.)
\footnote{An example: The bottom left cell in \fref{celllayout} 
with the coordinates 0/3 (bit pattern 000/011) 
has the number corresponding to the bit pattern 001010, i.\,e.~10. 
Incrementation by one yields the cell number 11.}

\begin{figure}
\setlength{\unitlength}{1.0cm}
\begin{center}
\begin{picture}(4,4)
\put(0,3){\makebox(1,1){1}}
\put(1,3){\makebox(1,1){2}}
\put(0,2){\makebox(1,1){3}}
\put(1,2){\makebox(1,1){4}}
\put(2,3){\makebox(1,1){5}}
\put(3,3){\makebox(1,1){6}}
\put(2,2){\makebox(1,1){7}}
\put(3,2){\makebox(1,1){8}}
\put(0,1){\makebox(1,1){9}}
\put(1,1){\makebox(1,1){10}}
\put(0,0){\makebox(1,1){11}}
\put(1,0){\makebox(1,1){12}}
\put(2,1){\makebox(1,1){13}}
\put(3,1){\makebox(1,1){14}}
\put(2,0){\makebox(1,1){15}}
\put(3,0){\makebox(1,1){16}}
\put(0,0){\framebox(4,4){}}
\put(0,2){\framebox(4,2){}}
\put(0,0){\framebox(2,2){}}
\put(2,1){\framebox(2,1){}}
\put(2,0){\framebox(2,1){}}
\end{picture}
\end{center}
\caption{Cell numbering layout in 2D with an example of domain decomposition. 
  Inhomogeneous load distribution leads to unequal domain sizes.}
\label{fig:celllayout}
\end{figure}

Then a block of $2^n$ consecutive cell numbers is affiliated with each process,
where $n$ can be different for each process.
Suppose the cells 1-16 in \fref{celllayout} should be distributed 
over 4 processes.
If a lot of particles are aggregated in the lower right corner, load-balancing 
could result in the following layout: Process I is affiliated with cells 1-8, 
process II with cells 9-12, process III with cells 13-14, 
and process IV with cells 15-16.

Every now and then the load of the processes has to be checked.  
A reasonable and simple measure of the load is 
the number of particles or the number 
of collisions in the process domain.
If a restructuring of the domain decomposition would result in
a significantly better load-balance, three processes, respectively, exchange 
information about their particle and cell data, so that two 
light-weight neighboring processes can merge their domains 
and a heavy-weight process can split its domain in two halves.
In the example above, if the initially inhomogeneous system 
becomes homogeneous, this procedure could result in the merging 
of the domains of processes III and IV and a splitting 
of the domain of process I. 
Process II does not participate in this rebalancing, but it should be informed
that its neighbor processes have changed, of course. 
In the end the layout would look like that: 
Process I is affiliated with cells 1-4, 
process II with cells 9-12, process III with cells 5-8, 
and process IV with cells 13-16.

\subsection{Causal Order}
\label{sec:causalorder}

A parallel approach to simulate asynchronous events has to make use of 
the concept of local times.
Each process handles the events in its domain and 
thereby increases the local simulation time.
When particles cross domain boundaries, 
the affected processes communicate with 
each other and events are inserted into or removed from the event lists 
of the processes.
If the event time of a newly inserted particle is less than 
the local simulation time, causality is violated, 
because a collision of that particle with another one, 
which has already been processed, could have been missed.

In general, there are two strategies to circumvent this problem:
In a {\em conservative} approach only those processes that are guaranteed 
not to violate causality are allowed to handle their events, the rest 
of the processes has to idle. 
In an {\em optimistic} approach the processes do not have to idle.
If causality is violated, a complex rollback protocol 
which corrects the erroneous computations has to be executed.
Whether a conservative approach is efficient or not depends highly on the
maximal event propagation speed.

Unfortunately, in event-driven molecular dynamics there is no upper 
limit for the speed of
events propagating along a chain of particles \cite{lubachevsky92},
even if the particles themselves are moving very slowly.
In other words, in the conservative case one process is operating and all 
others are idling and we are back to the serial algorithm.
So we are left with the optimistic strategy and have to undertake the task 
of implementing a rollback protocol.

\subsection{State Saving and Error Recovery}
\label{sec:statesave}
\label{sec:errorrecovery}

When a causality error is detected, the simulation is restarted from 
the latest saved state.
So the algorithm has to make a backup copy of the simulation data 
periodically and has to ensure that there is no latent causality error 
in this backup. 
The latter is guaranteed if all processes are synchronized at saving time.
This means that only those processes are allowed to continue 
their computations whose local simulation times have not yet reached 
synchronization time.
Note that there are other operations, like e.\,g.\ data output 
or load-balancing, which require periodic synchronization anyway.
Besides, without synchronization the local simulation times tend to
drift apart, which makes causality errors very likely \cite{korniss01}.

To find the optimal backup interval, we make use of an adaptive strategy.
If no causality error turns up between two successive save operations, 
the interval increases, otherwise it decreases.
If other operations trigger synchronization, this occasion is used for 
state saving, too, of course. 

If a causal error occurs, all processes perform a rollback to the saved state.
Furthermore a synchronization barrier is scheduled for the time 
when the error occured.
This prevents the same causality error from happening again.
Of course, another error could occur at an earlier point of time.
Then the simulation would perform another rollback and 
an earlier resynchronization would be scheduled.

For comparison the algorithm described in \cite{lubachevsky92} 
needs two backup copies of the simulation data.
So our strategy reduces memory consumption further by a factor $3/2$.

\subsection{Border Zone and Process Communication}
\label{sec:borderzone}
\label{sec:communication}
\label{sec:paroutline}

\begin{figure}
  \framebox[\boxlen]{
    \parbox{\pboxlen}{
      \begin{flushleft}
      {\bf parallel loop:}
      \begin{enumerate}
      \item communication about border zone events $\to$ timestep
      \item timestep $\gets$ min(timestep,time(synchronous tasks))
      \item for(timestep) {\bf event processing loop} (see \fref{serial}),\\ 
        send border zone particle information
      \item receive and process particle informations 
      \item error detection
      \item if(error) rollback
      \item if(global min(current time) = time(synchronous tasks))\\ 
        state saving, load-balancing, data output, \dots 
      \item goto 1
      \end{enumerate}
      \end{flushleft}
      }
    }
  \caption{Outline of the parallel algorithm}
  \label{fig:parallel}
\end{figure}

The border zone (in \cite{lubachevsky92} it is called {\em insulation layer}) 
consists of the cells $C_{\rm border}$ whose neighbors belong to a different 
process.
With the domain decomposition described in \sref{domain}, there 
is always a \lqq{}monolayer\rqq{} of border zone cells 
at the boundary of a process area. 

Each process thus maintains a list of virtual border zone cells 
which actually belong to its neighboring processes and 
a list of virtual particles residing in the virtual border zone cells.
Those virtual particles can act as partners for real particles. 
But no events are calculated for them directly and 
they are not represented in the event list, 
since they are real particles on another process.
The events are already calculated there and will be 
communicated to the adjacent processes.

However, it is highly inefficient for a process to communicate 
after every event with its neighbors.
So the parallel algorithm is designed in a stepwise manner 
(see \fref{parallel}):
Each computing step lasts until the next event in the border zone 
of a process, which is obtained in the following way:
An event is associated with each particle. 
Each event belongs to a cell 
(or two adjacent cells, if two colliding particles 
reside in different cells or if a particle changes from one cell to another).
The smallest event time in a cell is called the cell time $t_{\rm cell}$.
These cell times are stored in an additional priority queue 
similar to the event list in \sref{pqueue}.
This list contains the scheduled cell times for all local border zone cells 
and returns the minimum time 
$t_{\rm step}=\min_c{(t_{\rm cell}(c))}=t_{\rm cell}(c^\star)$ 
which belongs to cell $c^\star$.

Now, this time is used as the preliminary length of the calculation step.
But firstly, the neighboring cells of $c^\star$, if located on 
other processes are checked.
If they have scheduled an even earlier event, $t_{\rm step}$ will be shortened 
to that event.
The communication is done in the following way:
Each process sends $c^\star$ and $t_{\rm cell}(c^\star)$ as a query  
to the neighboring processes, which check the adjacent cells 
for their event times and reply with the minimum thereof.
If this answer is smaller, then it is used as $t_{\rm step}$ instead
(see \fref{parallel}, step 1).
Periodic tasks like data output, load-balancing or state saving, 
which require synchronization of the processes can shorten $t_{\rm step}$ 
even more (see \fref{parallel}, step 2).

One could think of more rigid policies to determine the length of a step,  
see e.\,g.\ \cite{lubachevsky92}.
We have also tried other strategies that reduce the number of rollbacks.
But on the other hand, they reduce parallelism, too.
A large part of the processes are idling, the communication overhead increases,
and the overall performance goes down.

After the communication phase, parallel event processing starts
and proceeds until the calculated point of time $t_{\rm step}$ 
(see \fref{parallel}, step 3), i.\,e.\ on average 
$\order{C/C_{\rm border}}$ iterations are processed.
If the algorithm encounters an earlier border zone event 
which has not been anticipated, the event processing step 
is stopped immediately to prevent the occurence of a causality error.
But normally, the last processed event is a regular border zone event. 
Then, after the last particle state update, the state information 
has to be communicated to the neighboring processes.

When a process has finished its computing step,
it reads the particle state messages that it has received during 
this step and adapts its data structures accordingly 
(see \fref{parallel}, step 4).
Real particles can only be affected as collision partners of virtual particles.
For a virtual particle there are several possibilities: 
A virtual particle can become a real particle by changing from one process 
to another, it can remain a virtual particle, but with a different position, 
velocity or cell number, 
or finally it can emerge on or disappear from the virtual particle list, 
because it enters or leaves the border zone, respectively.
If a virtual particle becomes real and newly calculated events for 
this particle violate causality, i.\,e.\ they are happening before 
the local simulation time, an error is signaled.
Moreover, the processes check the border zone cell times and compare them with 
the replies to their neighboring processes.
If the actual cell time is smaller than the reply (which corresponds to the 
current time of the neighbor process), 
a causality error has already happened with high probability, because a 
collision has been missed on the neighbor process, 
and an error is signaled, too (see \fref{parallel}, step 5).

If no error is detected, the simulation continues with the next step.
Otherwise a global rollback is performed (see \fref{parallel}, step 6) 
and the erroneous simulation step 
is restarted from the latest saved state (see \sref{statesave}).
To prevent the reappearance of the same error, a synchronization barrier 
is issued at the time when the error occured.

Other synchronization barriers can stem from periodic tasks such as 
state saving, load-balancing or data output (see \fref{parallel}, step 7).

However, most of all parallel loops do not include synchronization 
of the processes, but only communication between them.

\subsection{Parallel Performance}
\label{sec:performance}

First of all, parallelization imposes several performance penalties 
on the algorithm, among them
communication overhead, idle time, state saving, 
and rollbacks after erroneous computation steps.
On the other hand, the computational costs are shared among several 
processes.
If the latter outweighs the former, parallelization can be considered
succesful.
The same holds true for memory requirements.
State saving makes a copy of the whole system state and therefore doubles the 
memory usage.
But here as well parallelization combines the limited resources 
of single processes. 

\begin{figure}
  \begin{center}
    \includegraphics*[width=\gxlen]{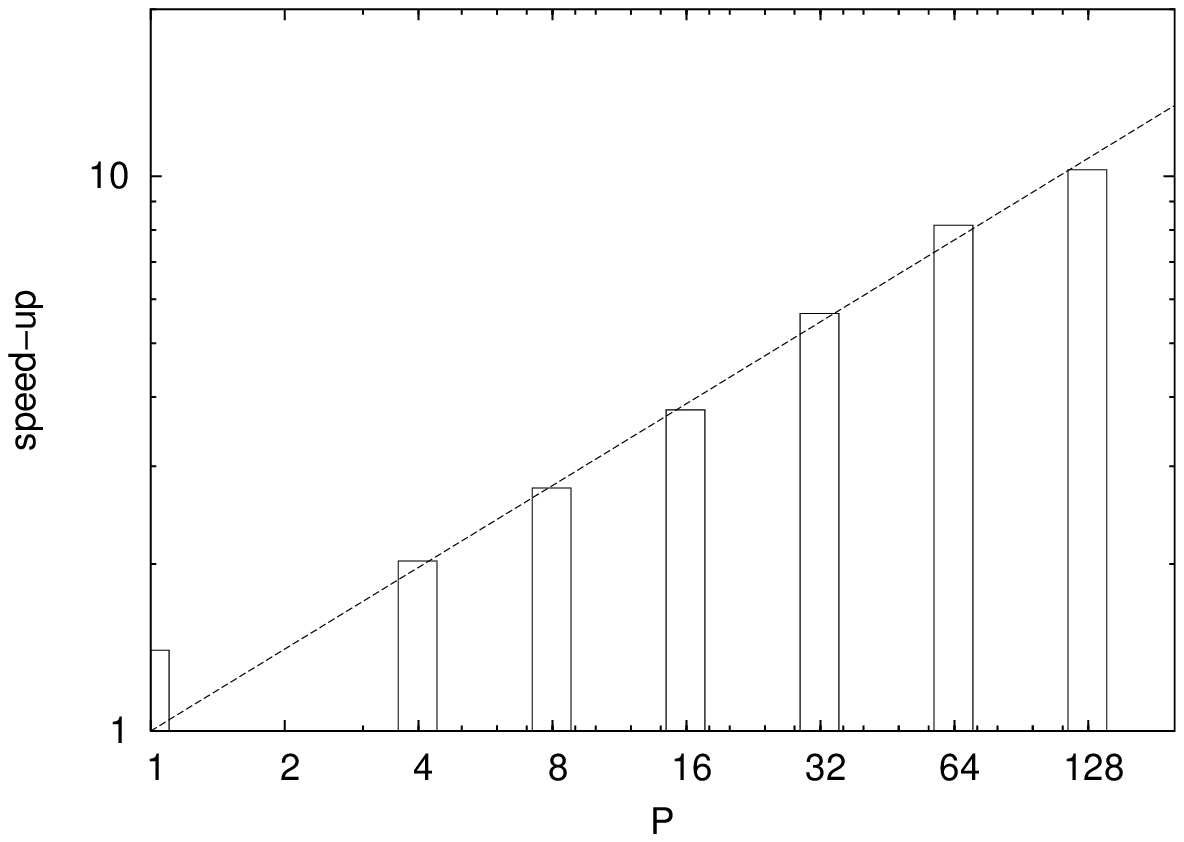}\\*
  \end{center}
    \caption{Speed-up for different numbers $P$ of processes in 2D. 
      $N = 5 \cdot 10^5$ particles, volume fraction $\nu = 0.3$, 
      $C = 1024^2 \approx 10^6$ cells. 
      The dashed line has a slope of 0.49.
      The data point for 1 process deviates from the dashed line, 
      because the serial algorithm has no communication overhead.
      Note the logarithmic axes.
      }
    \label{fig:par2d}
  \begin{center}
    \includegraphics*[width=\gxlen]{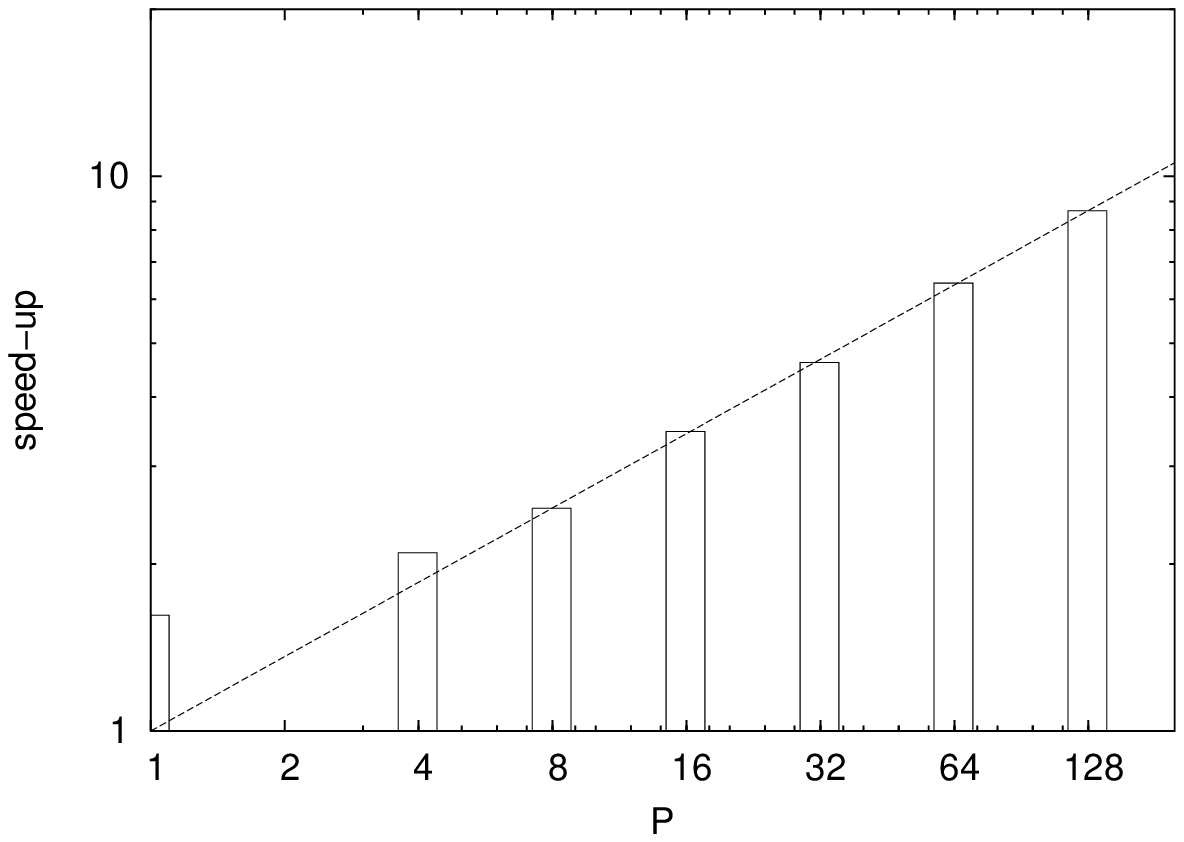}\\*
  \end{center}
    \caption{Speed-up for different numbers $P$ of processes in 3D. 
      $N = 2 \cdot 10^6$ particles, volume fraction $\nu = 0.25$, 
      $C = 128^3 \approx 2 \cdot 10^6$ cells. 
      The dashed line has a slope of 0.45.
      }
    \label{fig:par3d}
\end{figure}

Figs.\ \ref{fig:par2d}-\ref{fig:par3d} show that the speed-up of 
the parallelization (i.\,e. the reciprocal of the simulation time) 
for fixed system size is approximately 
proportional to $P^{1/2}$,
where $P$ is the number of processes.
\begin{figure}
  \begin{center}
    \includegraphics*[width=\gxlen]{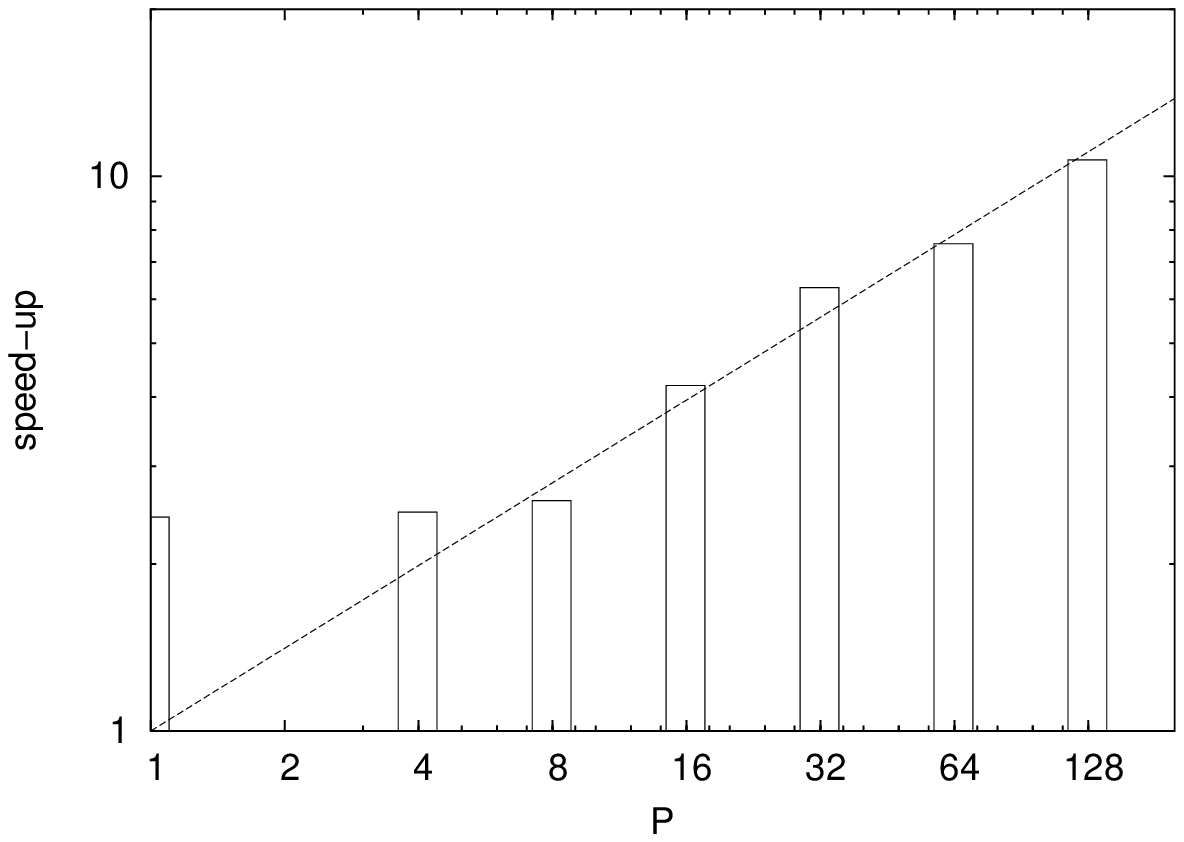}\\*
  \end{center}
    \caption{Speed-up for different numbers $P$ of processes in 3D. 
      $N/P = 5 \cdot 10^4$ particles per process, 
      volume fraction $\nu = 0.2$. 
      The dashed line has a slope of 0.50.
      }
    \label{fig:par3dc}
\end{figure}
Furthermore, as shown in \fref{par3dc}, the parallelization is also scalable 
if the number of processes is chosen proportional to the system size.
There are deviations for small $P$ when the effect of 
the parallelization overhead is large.
In addition, it has been shown in \cite{lubachevsky92} 
that the error recovery method 
presented in \sref{errorrecovery} is not scalable for $P \to \infty$.
But for practical purposes, at least until $P=128$, the algorithm remains 
scalable.

We have tested the parallel algorithm on a computer cluster with 
Pentium III 650 MHz processors and a 1.28 GBit/s switched LAN.
On a supercomputer with optimized communication hardware, scalability should be
even better.

For real physical applications with more than $10^2$ collisions per particle,
the maximal number of particles typically is limited by both 
available memory (see \fref{mem3d}) and computing time to about $10^6$ 
particles per process, i.\,e. a total number of particles of about $10^8$.

\begin{figure}
  \begin{center}
    \includegraphics*[width=\gxlen]{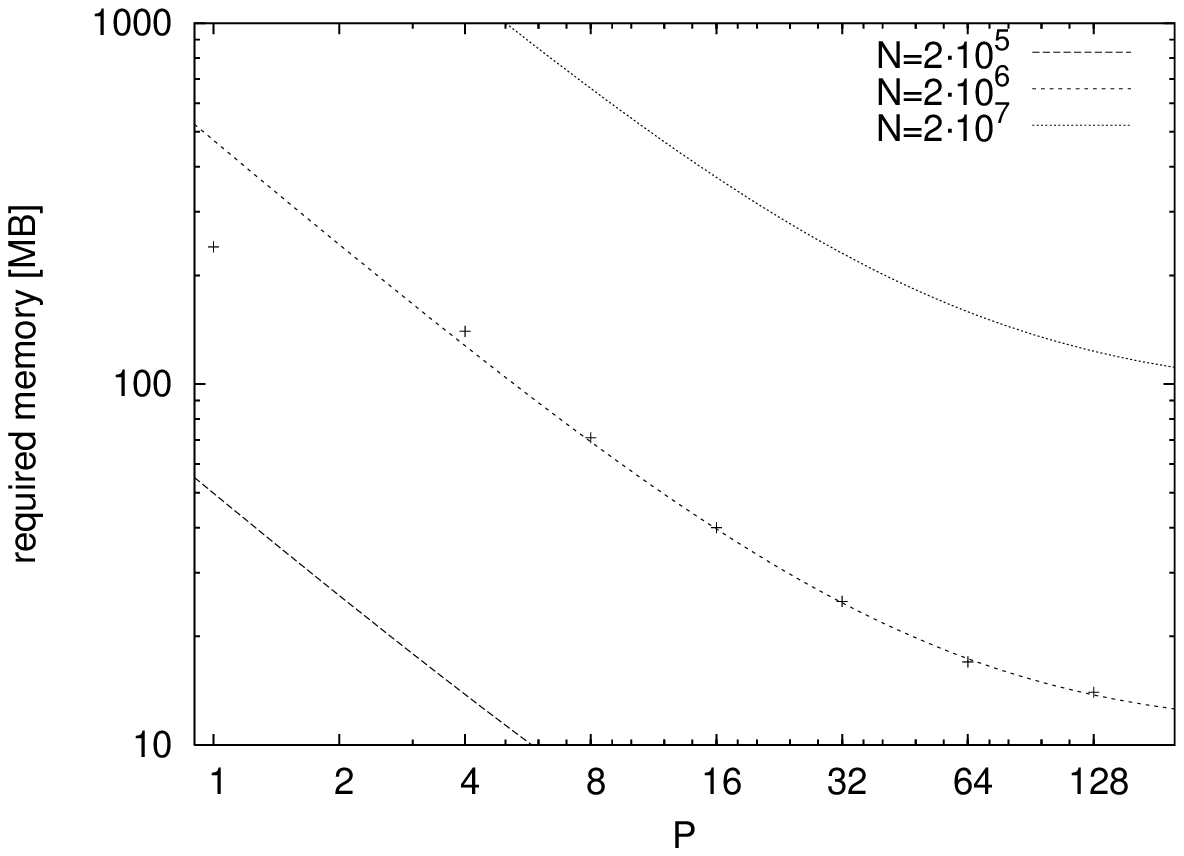}\\*
  \end{center}
    \caption{Memory requirements per process
      for different numbers $P$ of processes in 3D. 
      The points represent the measured values for $N = 2 \cdot 10^6$.
      The curves are obtained by a simple estimate 
      $c_1 N/P+c_2 (N/P)^{2/3}+c_3 N$, where the first
      term refers to real particles, the second term to virtual particles, and
      the third term to global data.
      The data point for 1 process deviates from the estimated value, 
      because the serial algorithm has no need for state saving.
      }
    \label{fig:mem3d}
\end{figure}

\section{Summary}
\label{sec:summary}

We have demonstrated how to parallelize event-driven molecular dynamics 
successfully.
Our algorithm is based on some ideas from \cite{lubachevsky92}, 
i.\,e.\ we have used an optimistic parallelization approach 
which performs a rollback protocol if a causality error occurs.
But the algorithm is enhanced in several ways:

Firstly, we have implemented dynamic load-balancing, which 
makes simulation of inhomogeneous systems possible.
Computing time is further reduced by an adaptive linked-cell structure 
which determines the optimal cell sizes.

Secondly, we have transferred the shared memory approach of 
\cite{lubachevsky92} to distributed memory architectures as well.
The parallelization has been realized with MPI.
In order to minimize idle time, we have made use of asynchronous 
communication, i.\,e.\ send and receive actions are decoupled from each other.
In addition, the amount of communication is limited to a minimum:
The event processing can continue steadily until a border zone event 
takes place on the local process or is expected to take place 
on a neighboring process. 
Only then the calculation has to be interrupted in order to communicate 
with the neigboring processes.
So, even on a cluster with rather poor communication hardware, parallelization 
yields a speed-up proportional to $P^{1/2}$\,---\,at least 
up to $P=128$ parallel processes. 

Thirdly, we have optimized the data structure of the algorithm.
With event-driven molecular dynamics insufficient memory 
is often a more serious problem than computing time.
In total our optimizations reduce memory requirements to one third
as compared to the method  proposed in \cite{lubachevsky92}.
This enabled us to perform simulations of real physical problems with
up to $10^8$ particles.

\bibliographystyle{unsrt}
\bibliography{bib/ed,bib/lui,bib/own}

\end{document}